\documentclass[mathleft]{an}
\usepackage{graphicx,bm}
\usepackage{times}
\renewcommand{\vec}[1]{\mbox{\boldmath$#1$}}
\def\gsim{\lower.4ex\hbox{$\;\buildrel >\over{\scriptstyle\sim}\;$}}
\def\lsim{\lower.4ex\hbox{$\;\buildrel <\over{\scriptstyle\sim}\;$}}

\newcommand{\kf}{k_{\rm f}}
\newcommand{\cp}{c_{\rm p}}
\newcommand{\cv}{c_{\rm v}}
\newcommand{\cs}{c_{\rm s}}
\newcommand{\csz}{c_{\rm s0}}
\newcommand{\kaptz}{\kappa_{\rm t0}}
\newcommand{\urms}{u_{\rm rms}}
\newcommand{\uu}{\bm{u}}
\newcommand{\dd}{{\rm d}}
\newcommand{\DD}{{\rm D}}

\newcommand{\nab}{\bm{\nabla}}

\begin{document}

\Pagespan{789}{}
\Yearpublication{2011}%
\Yearsubmission{2010}%
\Month{11}%
\Volume{999}%
\Issue{88}%

\title{Transport of angular momentum and chemical species by
    anisotropic mixing in stellar radiative interiors}

\author{L.L. Kitchatinov\inst{1,2,3}\fnmsep\thanks{Corresponding author:
  \email{kit@iszf.irk.ru}\newline}
\and  A. Brandenburg\inst{3,4} }

\titlerunning{Turbulent transport in stellar radiative cores}

\authorrunning{L.L. Kitchatinov \& A. Brandenburg}

\institute{Institute for Solar-Terrestrial Physics, P.O. Box 291,
Irkutsk 664033, Russia
 \and
Pulkovo Astronomical Observatory, St. Petersburg 196140, Russia
 \and
Nordita, AlbaNova University Center, Roslagstullsbacken 23, SE-10691
Stockholm, Sweden
 \and
Department of Astronomy, Stockholm University, SE-10691 Stockholm,
Sweden}

\received{} \accepted{} \publonline{}


\abstract{Small levels of turbulence can be present in stellar
radiative interiors due to, e.g., instability of rotational shear.
In this paper we estimate turbulent transport coefficients for
stably stratified rotating stellar radiation zones. Stable
stratification induces strong anisotropy with a very small ratio of
radial-to-horizontal turbulence intensities. Angular momentum is
transported mainly due to the correlation between azimuthal and radial
turbulent motions induced by the Coriolis force. This non-diffusive
transport known as the $\Lambda$-effect has outward direction in
radius and is much more efficient compared to the effect of radial
eddy viscosity. Chemical species are transported by small radial
diffusion only. This result is confirmed using direct numerical
simulations combined with the test-scalar method. As a consequence
of the non-diffusive transport of angular momentum, the estimated
characteristic time of rotational coupling (\lsim 100~Myr) between
radiative core and convective envelope in young solar-type stars is
much shorter compared to the time-scale of Lithium depletion ($\sim
1$~Gyr).
   }

\maketitle

\section{Introduction}
Mixing of material in stably stratified stellar radiative interiors
is opposed by buoyancy forces. Nevertheless, a certain level of
turbulence resulting, e.g., from instabilities of rotational shear
(Goldreich \& Schubert \cite{GS67}) or forcing from the upper
convection zone (Bl\"ocker et al.\ \cite{Bea98}) is possible and even
necessary to explain rotational coupling between the radiative core and
the convective envelope, and thus the depletion of light elements in solar-type
stars.

Turbulent transport in stellar radiation zones is supposed to be
strongly anisotropic (Spiegel \& Zahn \cite{SZ92}; Denissenkov
\cite{D10}). Buoyancy forces suppress radial mixing so that much
higher horizontal compared with radial velocities might be expected. The
degree of the anisotropy is not certain, however. It remains a
free parameter of transport models. The anisotropy in rotating
fluids is, however, not free (R\"udiger \& Pipin \cite{RP01}).
The influence of the Coriolis force on horizontal motions produces
radial displacements. The ratio of radial to azimuthal mixing
intensities is controlled by the balance between Coriolis and
buoyancy forces.

The aim of this paper is to estimate the effects of rotation and
stable stratification on turbulence. We do not specify here the origin of
the turbulence but just prescribe the so-called \lq original turbulence'
that would take place in non-rotating and neutrally stratified
fluid. The combined effect of rotation and stable
stratification on this original turbulence is then estimated.

We shall see that the resulting parameters of turbulent mixing
depend only slightly on the properties of the prescribed original
turbulence. In particular, the anisotropy,
 \begin{equation}
    \frac{\langle u_\phi^2\rangle}{\langle u_r^2\rangle} \approx
    \frac{\tau^2 N^4}{\Omega^2},
 \label{1}
 \end{equation}
varies only slightly between very different prescriptions (in this
equation, $\vec{u}$ is the fluctuating velocity, $\Omega$ is the angular
velocity, $N$ is the buoyancy frequency, and $\tau$ is the eddy turnover
time). Typically, $\Omega \ll N$ in radiation zones and probably
we also have $\tau N \gg 1$.
Therefore, the anisotropy of Eq.~(\ref{1}) is
high. Our main finding is that the strongly anisotropic turbulence
is much more efficient in transporting angular momentum than
chemical species. The reason is that the Coriolis force makes
the azimuthal and radial motions correlated. Fluid particles
with positive (negative) azimuthal velocity are pushed by the
Coriolis force in the positive (negative) radial direction. As a
result, $\langle u_ru_\phi\rangle
> 0$, and the turbulence transports angular momentum in the positive
radial direction. This angular momentum flux cannot be interpreted
as an effect of eddy viscosity, i.e., it is not diffusive by nature.
Transport of chemical species, on the contrary, is only due to
radial turbulent mixing.

The non-diffusive transport of angular momentum that we find is
essentially the well-known $\Lambda$-effect of differential rotation
theory (Lebedinskii \cite{L41}; R\"udiger \cite{R89}). The effect
has been well studied for nearly adiabatically stratified stellar
convection zones. Our paper suggests that it can be important for
the rotational coupling between radiative interiors and convective
envelopes as well.

\section{Mathematical formulation}
\subsection{Relation between the original and background turbulence}
To specify the effects of rotation and stable stratification on
turbulence, we follow the standard approach of quasilinear theory by
formulating the linear relation, $u_i = D_{ij}u^{(0)}_j$ between the
velocity fields of original turbulence ${\vec u}^{(0)}$, which
would take place in nonrotating neutrally stratified fluid, and
actual or background turbulence $\vec u$. The relation tensor
$D_{ij}$ includes the effects of rotation and stratification.

We apply a simple version of the $\tau$-approximation to write the
equations for fluctuating velocity and entropy as follows,
\begin{eqnarray}
    \frac{\vec u}{\tau} &+&
    2{\vec\Omega}\times{\vec u} +
    \frac{1}{\rho}{\vec\nabla}P  -
    \vec{g} = {\vec f} ,
    \nonumber \\
    \frac{s'}{\tau}  &=& - ({\vec u}\cdot{\vec\nabla})\langle s\rangle
    ,
    \label{2}
\end{eqnarray}
where $\vec f$ is a random force driving the turbulence, $\tau$ is the
eddy turnover time, $\vec{g}$ is gravity and the $\tau$-relaxation
terms replace the nonlinear terms together with time-derivatives,
\begin{eqnarray}
    &&\frac{\partial\vec u}{\partial t} + ({\vec u}\cdot{\vec\nabla})
    {\vec u} - \langle({\vec u}\cdot{\vec\nabla})
    {\vec u}\rangle  \rightarrow \frac{\vec u}{\tau} ,
    \nonumber \\
    &&\frac{\partial s'}{\partial t} + {\vec u}\cdot{\vec\nabla}s -
    \langle {\vec u}\cdot{\vec\nabla}s\rangle \rightarrow
    \frac{s'}{\tau}.
    \label{3}
\end{eqnarray}
Microscopic diffusion is neglected. The $\tau$-approximation
(\ref{3}) assumes that the nonlinear interaction of turbulent eddies
leads to their effective dissipation (a turbulent fragmentation of
scales) in a characteristic time $\tau \approx \ell /u$, where
$\ell$ is the characteristic spatial scale of the eddies. We assume
that the motion is incompressible, $\mathrm{div}{\vec u} = 0$. This
means that vertical displacements are small compared to the density
scale height and velocities are small compared to the speed of
sound.

Fourier transformation is applied,
\begin{eqnarray}
    {\vec u}({\vec r}) &=& \int \mathrm{e}^{\mathrm{i}{\bm k}\cdot{\bm r}}
    \tilde{\vec u}({\vec k})\ \mathrm{d}{\vec k} ,
    \nonumber \\
    s'({\vec r}) &=& \int \mathrm{e}^{\mathrm{i}{\bm k}\cdot{\bm r}}
    \tilde{s}'({\vec k})\ \mathrm{d}{\vec k},
    \label{4}
\end{eqnarray}
to convert the partial differential equations (\ref{2}) into
algebraic equations. We neglect variations of mean fields on the
spatial scale of the turbulence to obtain a closed equation for
the fluctuating velocity:
\begin{equation}
    \tilde{u}_i + {N^*}^2( \hat{r}_i -
    \mu \hat{k}_i)(\hat{\vec{r}}\cdot\tilde{\vec u}) + \sigma \Omega^* \varepsilon_{ipm}
    \hat{k}_p\tilde{u}_m = \tilde{u}^{(0)}_i
    \label{6}
\end{equation}
(cf.\ Kitchatinov \& R\"udiger (\cite{KR08}) on how the pressure
term is treated). In Eq.~(\ref{6}), $\hat{\vec r}$ is the radial
unit vector, $\hat{\vec k} = {\vec k}/k$ is the unit vector along
the wave vector, $\mu = (\hat{\vec k}\cdot\hat{\vec r})$ is the cosine
of the angle between wave vector and radius, and $\sigma =
(\hat{\vec k}\cdot{\vec\Omega})/\Omega$ is cosine of the angle
between the wave vector and rotation axis. The Coriolis number
\begin{equation}
    \Omega^* = 2\tau\Omega
    \label{7}
\end{equation}
and the normalized buoyancy frequency
\begin{equation}
    N^* = \tau N,\ \ \ \ \ \
    N^2 = \frac{g}{c_\mathrm{p}}\frac{\partial\langle
    s\rangle}{\partial r} ,
    \label{8}
\end{equation}
parameterize the effects of rotation and stratification.

Solving Eq.~(\ref{6}) for $\tilde{\vec u}$ gives the relation
tensor,
\begin{eqnarray}
    D_{ij} &=& \frac{1}{1 + {N^*}^2 (1 - \mu^2) +
    \sigma^2{\Omega^*}^2} \times
    \nonumber \\
    && \bigg( \left( 1 + {N^*}^2 (1-\mu^2)\right) \left( \delta_{ij}
    + \sigma\Omega^*\varepsilon_{ijp}\hat{k}_p\right) -
    \nonumber \\
    &&-\ {N^*}^2 (\hat{r}_i - \mu \hat{k}_i)\hat{r}_j +
    \nonumber \\
    &&+\ \sigma \Omega^*{N^*}^2\left( (\hat{r}_i - \mu
    \hat{k}_i)\varepsilon_{jmp} - \hat{r}_j\varepsilon_{imp}\right)
    \hat{r}_m\hat{k}_p\bigg) ,
    \nonumber \\
    \tilde{u}_i &=& D_{ij}\tilde{u}_j^{(0)} .
    \label{9}
\end{eqnarray}
This equation describes the joint influence of rotation and stable
stratification on the turbulence.

In the limit of neutral stratification, $N^* \rightarrow 0$,
Eq.~(\ref{9}) reduces to the familiar expression (Kitchatinov \cite{K86})
\begin{equation}
    D_{ij} = \frac{\delta_{ij}
    + \sigma\Omega^*\varepsilon_{ijp}\hat{k}_p}{1 +
    \sigma^2{\Omega^*}^2},
    \label{10}
\end{equation}
describing the influence of rotation in the $\tau$-approximation.
In the other limit of slow rotation, $\Omega^* \rightarrow 0$,
Eq.~(\ref{9}) reduces to
\begin{equation}
    D_{ij} = \delta_{ij} - \frac{{N^*}^2}{1 + {N^*}^2 (1-\mu^2)}
    (\hat{r}_i - \mu \hat{k}_i)\hat{r}_j .
    \label{11}
\end{equation}
This equation accounts for the effect of stable stratification
alone.
\subsection{Original turbulence models}
The original turbulence properties can be prescribed by specifying
the spectral tensor $\tilde{Q}^{(0)}_{ij}$:
\begin{equation}
    \langle \tilde{u}^{(0)}_i({\vec k})\tilde{u}^{(0)}_j({\vec k}')\rangle =
    \tilde{Q}^{(0)}_{ij} ({\vec k})\ \delta ({\vec k} + {\vec k}')
    \label{12}
\end{equation}
(R\"udiger \cite{R89}). The relation tensor (\ref{9}) can then be
used to account for the effects of rotation and stratification,
\begin{eqnarray}
    \tilde{Q}_{ij} &=& D_{im}D_{jn}\tilde{Q}^{(0)}_{mn} ,
    \nonumber \\
    Q_{ij} &=& \langle u_iu_j\rangle = \int \tilde{Q}_{ij}({\vec k})\
    \mathrm{d}{\vec k} .
    \label{13}
\end{eqnarray}

For the simplest case of isotropic nonhelical turbulence the spectral tensor
reads
\begin{equation}
    \tilde{Q}_{ij}^{(0)} = \frac{E(k)}{8\pi k^2}\left( \delta_{ij} -
    \hat{k}_i \hat{k}_j\right) .
    \label{14}
\end{equation}
In this equation, $E$ is the spectrum function
\begin{equation}
    \langle u^2\rangle^{(0)} = \int\limits_0^\infty E(k) \mathrm{d}
    k .
    \label{15}
\end{equation}

A more realistic model is anisotropic turbulence with a preferred
direction being the radial one.
Horizontal velocities are expected to be much
larger than the vertical ones. For the extreme case of strictly
horizontal random motions, we have
\begin{eqnarray}
    \tilde{Q}_{ij}^{(0)} = q(k,\mu ) \times
    \ \ \ \ \ \ \ \ \ \ \ \ &&
    \nonumber \\
    \big( (1-\mu^2)(\delta_{ij} - \hat{k}_i
    \hat{k}_j) &-& (\hat{r}_i-\mu \hat{k}_i)(\hat{r}_j - \mu \hat{k}_j)\big) .
    \label{16}
\end{eqnarray}
The correlation lengths for vertical and horizontal directions may differ.
If, however, the lengths are equal, the $q$-function in
Eq.~(\ref{16}) does not depend on $\mu$:
\begin{equation}
    q = \frac{3E(k)}{8\pi k^2} .
    \label{17}
\end{equation}
The case where the correlation length in radius is small compared to
the horizontal correlation length can be modeled by the equation
\begin{equation}
    q = \frac{3 E(k_\perp)\ k_\perp}{8\pi k^3},\ \ \ \ \ \
    k_\perp = k\sqrt{1 - \mu^2} .
    \label{18}
\end{equation}
The spectrum functions $E$ of Eqs.~(\ref{17}) and (\ref{18}) are
related to the turbulence intensity by the same equation (\ref{15}) as
in the case of the isotropic turbulence of Eq.~(\ref{14}).

Anisotropic turbulence that has finite radial velocities (but
different from the horizontal velocities) can be modeled by a linear
superpositions of the spectral tensors (\ref{14}) and (\ref{16}). We
shall see that, if the buoyancy frequency is large compared to
the rotation frequency, $N \gg \Omega$, and the normalized buoyancy
frequency is large, $N^* \gg 1$, which are conditions typical of
stellar radiation zones, then the turbulent transport parameters
differ little between the representations (\ref{14}), (\ref{17}) and
(\ref{18}) for the original turbulence, i.e., the transport
characteristics are not sensitive to a particular choice of the
original turbulence model.

\subsection{Direct numerical simulations}
\label{DNS}

An independent verification of the effects of strong stratification on
turbulent diffusion is provided by means of direct numerical simulations.
In that case, we solve the full set of compressible hydrodynamic equations
for $\rho$, $\uu$, and $s$:
\begin{eqnarray}
\frac{\DD\rho}{\DD t}&\!=\!&-\rho\nab\cdot\uu,
\label{density}\\
\rho\frac{\DD\uu}{\DD t}&\!=\!&-\nab P
+\nab\cdot(2\rho\nu\bm{\mathsf{S}}) + \rho({\bm f}+{\bm g}),
\label{velocity}\\
\rho T\frac{\DD s}{\DD t} &\!=\!& \nab\cdot K \nab T
+ 2\rho\nu\bm{\mathsf{S}}^2-\frac{1}{\tau_s}\rho (\cs^2-\csz^2),
\label{entropy}
\end{eqnarray}
where $\DD/\DD t=\partial/\partial t+\uu\cdot\nab$ is the advective
derivative, ${\sf S}_{ij}=\textstyle{\frac{1}{2}}(u_{i,j}+u_{j,i})
-\textstyle{\frac{1}{3}}\delta_{ij}\nab\cdot\uu$ is the traceless
rate of strain tensor, commas indicate partial differentiation,
$\nu$ is the kinematic viscosity, the specific entropy is given by
$s=\cv\ln P-\cv\ln\rho$, where $\cp$ and $\cv$ are the specific
heats at constant pressure and constant volume, respectively, the
temperature is related to $P$ and $\rho$ via $(\cp-\cv)T=P/\rho$,
which, in turn, is related to the sound speed $\cs$ via
$\cs^2=\gamma P/\rho$, where $\gamma = \cp / \cv$ is the ratio of
specific heats, ${\bm f}$ is the external forcing function, and $K$ is the
thermal conductivity. The last term in the entropy equation
(\ref{entropy}) represents a cooling term that keeps the temperature
(or sound speed) approximately constant with a given relaxation or
cooling time $\tau_s$. The flow is driven by a random forcing
function consisting of nonhelical waves with wavenumbers whose
modulus lie in a narrow band around an average wavenumber $k_{\rm
f}$ (Haugen et al.\ \cite{HBD04}). We arrange the amplitude of the
forcing function such that the Mach number based on the rms velocity
remains below 0.1, so the effects of compressibility are negligible.

We consider a cubic domain of size $L^3$ with periodic boundary
conditions in the $x$ and $y$ directions and insulating impenetrable
stress-free boundary conditions on $z=\pm L/2$. The mass in the
volume is therefore conserved and given by $\int\rho\,\dd^3x=\rho_0
L^3$. The lowest wavenumber that fits into the domain is
$k_1=2\pi/L$. Our average forcing $\kf$ wavenumber is chosen such
that $\kf/k_1=5$. We adopt isothermal initial conditions with
$\cs=\csz$ and $\rho=\rho_0\exp(-z/H)$, where $H=\csz^2/\gamma g$
is the scale height which is chosen such that $H k_1=1$. The turnover
time based on the wavenumber $\kf$ is $\tau=(\urms\kf)^{-1}$, where
$\urms$ is the rms velocity based on all three velocity components.
We vary the normalized buoyancy frequency by varying the forcing
amplitude. For each run the viscosity is adjusted such that the
Reynolds number, $\mbox{Re}=\urms/\nu\kf$, is around 40. The Prandtl
number, $\mbox{Pr}=\nu\cp\rho/K$, is chosen to be equal to unity.
The cooling time is chosen such that $\tau_s\urms\kf$ is also around
unity.

To quantify the suppression of vertical mixing in a numerical
simulation of stratified turbulence we use the test-scalar method
by solving equations for the fluctuating passive scalar
concentration in the presence of a prescribed mean passive
scalar concentration,
\begin{equation}
{\partial c^{pq}\over\partial t}=-\nab\cdot(
\uu C^{pq}+\uu c^{pq}-\langle\uu c^{pq}\rangle)
+\kappa\nabla^2c^{pq},
\label{dcpqdt}
\end{equation}
where $\uu$ is the velocity fluctuation obtained from Eq.~(\ref{velocity}),
and the mean flow is zero.
The superscript $p$ (=1 or 3) stands for test scalars varying in the $x$
or $z$ directions while $q$ stands for c and s that represent mean
fields that are proportional to $\cos k x_p$ or $\sin k x_p$, respectively,
\begin{equation}
C^{pc}=C_0\cos kx_p,\quad C^{ps}=C_0\sin kx_p,\\
\label{TestScalar}
\end{equation}
where $(x_1,x_2,x_3)=(x,y,z)$ are Cartesian coordinates and
$C_0$ is a normalization factor.
Angle brackets denote planar averaging over the $yz$ plane (if $p=1$)
or the $xy$ plane (if $p=3$).
The passive scalar flux is then related to the gradient of $C^{pc}$ via
$\langle u_i c^{pq}\rangle=-\kappa_{ij}\nabla_j C^{pc}$,
where $\kappa_{ij}$ are the components of the turbulent diffusivity tensor,
obtained as
\begin{eqnarray}
\kappa_{xx}&\!=\!&(\sin kx\langle u_x c^{1c}\rangle-
\cos kx\langle u_x c^{1s}\rangle)/kC_0,
\\
\kappa_{zz}&\!=\!&(\sin kz\langle u_z c^{1c}\rangle-
\cos kz\langle u_z c^{1s}\rangle)/kC_0,
\end{eqnarray}
by using the solutions for four different test scalars.
These coefficients depend on time and one spatial coordinate,
but because of stationarity and approximate homogeneity of
the turbulence intensity, we present in the following temporal
and spatial averages of these coefficients.

The test scalar method has been used previously to quantify mixing
in turbulence in the presence of rotation and magnetic fields
(Brandenburg et al.\ \cite{BSV09}), shear (Madarassy \& Brandenburg
\cite{MB10}), as well as isothermal density stratification
(Brandenburg et al.\ \cite{BRK12}). However, unlike those earlier
works, the entropy equation is here included and a non-isothermal
equation of state for a perfect monatomic gas is used with $\gamma
=5/3$. Equations (\ref{density})--(\ref{dcpqdt}) are solved using
the {\sc Pencil
Code}\footnote{http://www.pencil-code.googlecode.com} in three
dimensions.
The numerical resolution used for the simulations presented
here is $128^3$ meshpoints.
Like the Prandtl, the Schmidt number, $\mbox{Sc}=\nu/\kappa$, is also
chosen to be equal to unity.

\section{The effect of stable stratification}
We consider first the effect of stable stratification for a
non-rotating fluid, $\Omega = 0$. The only preferred direction in
this case is the radial one. For any model of the original
turbulence, we therefore have
\begin{equation}
    \langle u_i u_j\rangle = \frac{\langle
    u^2_\mathrm{h}\rangle}{2}\left(\delta_{ij} - \hat{r}_i\hat{r}_j\right) +
    \langle u^2_r\rangle \hat{r}_i\hat{r}_j ,
    \label{19}
\end{equation}
where $\langle u^2_\mathrm{h}\rangle = \langle u^2_\phi\rangle +
\langle u^2_\theta\rangle$ is the horizontal turbulence intensity.
Standard spherical coordinates ($r,\theta ,\phi$) are used.

\begin{figure}
    \includegraphics[width=8.cm]{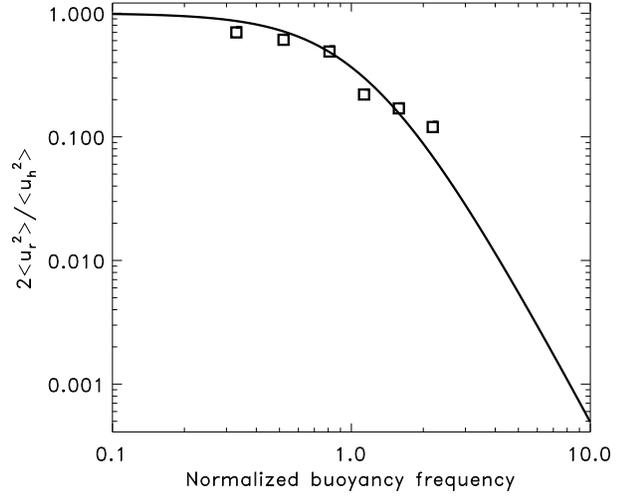}
    \caption{Ratio of the vertical to horizontal turbulence
    intensities in dependence on the normalized buoyancy frequency
    (\ref{8}) for the case of a non-rotating fluid and isotropic
    original turbulence. Squares show the results of 3D direct
    numerical simulations for isotropically forced turbulence.
              }
    \label{f1}
\end{figure}

On using the relation tensor (\ref{11}) and Eq.~(\ref{13}) for the
isotropic original turbulence of Eq.~(\ref{14}), we find
\begin{eqnarray}
    &&\langle u^2_\mathrm{h}\rangle = \langle u^2\rangle^{(0)}
    \bigg( \frac{1}{2} + \frac{1}{4{N^*}^2} -
    \nonumber \\
    && \left. - \frac{1}{8{N^*}^3\sqrt{{N^*}^2 + 1}}\ \mathrm{ln}
    \left(\frac{\sqrt{{N^*}^2 +
    1} + N^*}{\sqrt{{N^*}^2 + 1} - N^*}\right)\right) ,
    \label{20}
\end{eqnarray}
\begin{eqnarray}
    &&\langle u^2_r\rangle = \frac{\langle
    u^2\rangle^{(0)}}{4{N^*}^2 ({N^*}^2 +1)}\times
    \nonumber \\
    &&\left( \frac{2{N^*}^2 +
    1}{2N^*\sqrt{{N^*}^2 + 1}}\ \mathrm{ln}\left(\frac{\sqrt{{N^*}^2 +
    1} + N^*}{\sqrt{{N^*}^2 + 1} - N^*}\right) - 1\right) .
    \label{21}
\end{eqnarray}
Figure~\ref{f1} shows the ratio of vertical to horizontal turbulence
intensities as a function of the normalized buoyancy frequency
$N^*$. Squares on the plot show the results of direct numerical
simulations described in Sect.~\ref{DNS}.

Horizontal turbulence intensity is influenced by the stable
stratification only slightly. In the limit of very large $N^*$, it
is $\langle u^2_\mathrm{h}\rangle = \langle u^2\rangle^{(0)}/2$,
i.e., $\langle u^2_\mathrm{h}\rangle$ is reduced compared to the
case of a non-stratified fluid by a factor of 3/4 only. The vertical
motions, by contrast, are strongly suppressed:
\begin{equation}
    \langle u^2_r\rangle \simeq \frac{\langle
    u^2\rangle^{(0)}}{4{N^*}^4}\left(\mathrm{ln}(4{N^*}^2) -
    1\right)\ \ \ \mathrm{for}\ \ N^* \gg 1
    \label{22}
\end{equation}
in the strong stratification limit. The originally isotropic
turbulence is changed towards horizontal turbulence as $N^*$
increases.

As might be expected, the horizontal turbulence of Eq. (\ref{16}) is
not influenced by the stratification at all, i.e., $Q_{ij} =
Q^{(0)}_{ij}$ in this case. Therefore, whatever model for the
original turbulence in Section~2.2 is used, the resulting turbulence
in the limit of large $N^*$ is almost the same. The only difference
between isotropic and horizontal original turbulence models is
that for the first case the radial velocities of Eq.~(\ref{22}) are
finite, although very small.

\begin{figure}
\includegraphics[width=8.cm]{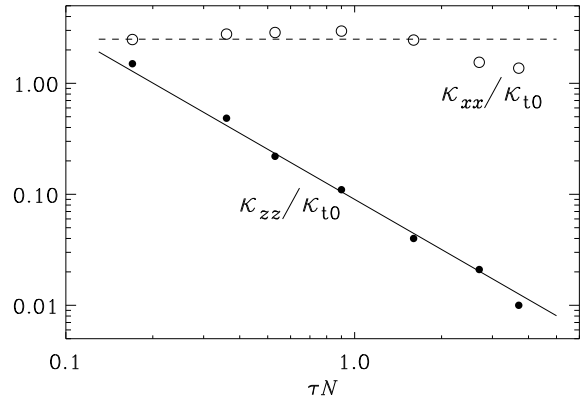}
\caption{Dependence of $\kappa_{xx}/\kaptz$ (open symbols) and
$\kappa_{zz}/\kaptz$ (filled symbols) on the normalized buoyancy frequency.
The dashed line shows that $\kappa_{xx}/\kaptz\approx2.5$ while the
solid line gives $\kappa_{zz}/\kaptz\approx0.09\,(\tau N)^{-3/2}$.
}
\label{pkap}
\end{figure}

\begin{figure*}
\includegraphics[width=\textwidth]{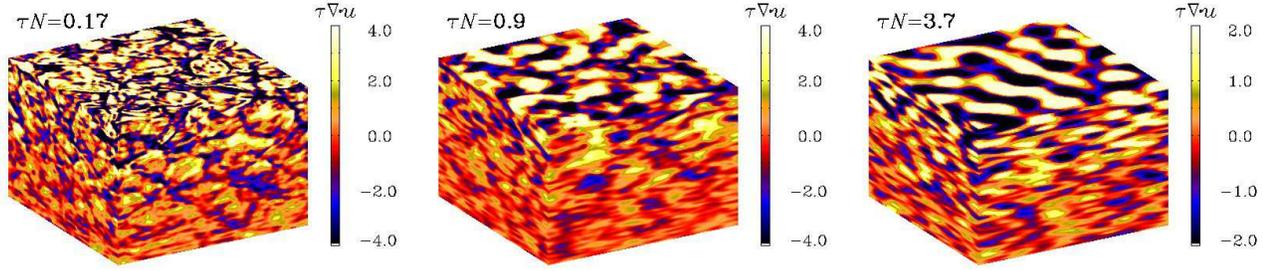}
\caption{Visualizations of $\tau\nab\cdot\uu$ on the periphery of the
computational domain for $\tau N=0.17$, 0.9, and 3.7.
}
\label{divu}
\end{figure*}

Similar results are obtained numerically for the turbulent diffusivity
using the test-scalar method; see Fig.~\ref{pkap}, where we plot $\kappa_{xx}$
and $\kappa_{zz}$, both normalized by the reference value $\kaptz=\urms/3\kf$.
Note that $\kappa_{xx}/\kaptz\approx2.5$, independent of $\tau N$,
while $\kappa_{zz}/\kaptz\approx0.09\,(\tau N)^{-3/2}$.
Visualizations of the scalar quantity $\tau\nab\cdot\uu$ on the periphery
of the computational domain (Fig.~\ref{divu}) confirm our expectation
that turbulent structures that become flatter as $\tau N$ is increased.
These structures become smoother in the $xy$ plane, while their length
scale in the $z$ direction decreases.

\section{Mixing in rotating and stratified fluids}
Similarity of the results for different original turbulence models
is even more pronounced when rotation is included. Coriolis force
produces radial motion even for original horizontal turbulence.
The relative intensity of radial mixing is controlled by the balance
between buoyancy and Coriolis forces.

Producing analytical results for arbitrary values of $N^*$ and
$\Omega^*$ is problematic. Derivations were performed for the
practically interesting case of strong stratification, $N^* \gg 1$,
and not too fast rotation, $\Omega^* \ll N^*$. The period of gravity
waves for radiation zones of solar-type stars is of the order of one
hour. The characteristic time of turbulence is probably much longer (it
is about one month for convection and probably longer for radiation
zone motions), so that the condition $N^* \gg 1$ is well satisfied. The
condition $\Omega^* \ll N^*$ should be satisfied as well in
radiation zones of pressure supported stars (otherwise we are
dealing with centrifugally supported disks).

We now use Eq.~(\ref{9}) for the relation tensor that includes the
combined effects of rotation and stratification. For the case of the
isotropic original turbulence of Eq.~(\ref{14}), the terms of lowest
order in ${N^*}^{-1}$ in the most significant velocity correlations
read
\begin{eqnarray}
    \langle u_\mathrm{h}^2\rangle &=& \frac{1}{2} \langle
    u^2\rangle^{(0)} ,
    \nonumber \\
    \langle u^2_r\rangle  &=& \frac{\langle u^2\rangle^{(0)}}{4{N^*}^4}
    \left[ \mathrm{ln}\left(\frac{4{N^*}^2}{1 +
    {\Omega^*}^2\cos^2\theta}\right) - 1 + {\Omega^*}^2 \right.
    \nonumber \\
    &+& \left. {\Omega^*}^2\cos^2\theta \left(\mathrm{ln}\left(\frac{4{N^*}^2}{1 +
    {\Omega^*}^2\cos^2\theta}\right) - 4\right)\right] ,
    \nonumber \\
    \langle u_r u_\theta\rangle &=& \langle u^2\rangle^{(0)}
    \frac{{\Omega^*}^2}{4{N^*}^4} \left(\mathrm{ln}\left(\frac{4{N^*}^2}{1 +
    {\Omega^*}^2\cos^2\theta}\right) - 3\right)
    \nonumber \\
    && \times\sin\theta\cos\theta ,
    \nonumber \\
    \langle u_r u_\phi\rangle &=& \langle u^2\rangle^{(0)}
    \frac{\Omega^*}{4{N^*}^2}\sin\theta .
    \label{23}
\end{eqnarray}
The turbulence intensities $\langle u^2_\mathrm{h}\rangle$ and
$\langle u_r^2\rangle$ control the eddy diffusion in latitude and
radius, respectively. The cross-correlation $\langle
u_ru_\phi\rangle$ is important for transport of angular momentum.
The correlation $\langle u_r u_\theta\rangle$ may cause the
temperature variation with latitude; positive $\langle u_r
u_\theta\rangle\cos\theta$ implies poleward eddy heat flux. It may
be noted that rotation does not produce anisotropy in the horizontal
plane in the strong stratification limit, $\langle u_\theta^2\rangle
= \langle u_\phi^2\rangle = \langle u_\mathrm{h}^2\rangle/2$ (more
precisely, the anisotropy is small: $(\langle u_\theta^2\rangle -
\langle u_\phi^2\rangle)/\langle u_\mathrm{h}^2\rangle \sim
{\Omega^*}^2/{N^*}^4$).

The results for horizontal turbulence of Eq.~(\ref{16}) with an
isotropic correlation length of Eq.~(\ref{17}) or a short vertical
correlation length of Eq.~(\ref{18}) are the same and read
\begin{eqnarray}
    \langle u_\mathrm{h}^2\rangle &=& \langle u^2\rangle^{(0)} ,
    \nonumber \\
    \langle u^2_r\rangle  &=& \langle u^2\rangle^{(0)}
    \frac{{\Omega^*}^2}{2{N^*}^4} ,
    \nonumber \\
    \langle u_r u_\theta\rangle &=& \langle u^2\rangle^{(0)}
    \frac{{\Omega^*}^2}{2{N^*}^4} \sin\theta\cos\theta ,
    \nonumber \\
    \langle u_r u_\phi\rangle &=& \langle u^2\rangle^{(0)}
    \frac{\Omega^*}{2{N^*}^2}\sin\theta .
    \label{24}
\end{eqnarray}
If the Coriolis number (\ref{7}) is not small ($\Omega^* > 1$),
equations (\ref{23}) and (\ref{24}), though different in details,
are of the same order of magnitude. As the $\tau$-approximation, which
was used to derive these equation, probably has the same accuracy, the
results are practically the same. We focus in the
following discussion on the results of Eq.~(\ref{24}) because the case of
horizontal original turbulence is probably more adequate for stably
stratified fluids.

\section{Discussion}
A remarkable feature of Eq.~(\ref{24}) is the finite
cross-correlation $\langle u_ru_\phi\rangle$. This means
that the turbulence transports angular momentum in radius. The
cross-correlation is positive so that the angular momentum is
transported outward. For the case of differential rotation caused by
stellar spin-down, this angular momentum flux acts in same direction
as turbulent viscosity, but it is not viscous by nature.

\begin{figure}
    \centerline{
    \includegraphics[width=5.cm]{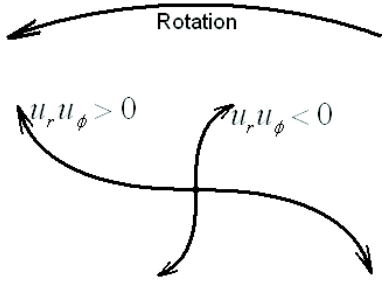}
    }
    \caption{Illustration of the origin of non-viscous angular
    momentum transport by anisotropic turbulence (see text).
              }
    \label{f2}
\end{figure}

What we find here is the known $\Lambda$-effect of differential
rotation theory (Lebedinskii \cite{L41}; R\"udiger \cite{R89}).
Figure~\ref{f2} illustrates its origin. The positive correlation
$\langle u_ru_\phi\rangle$ results from the Coriolis force action on
the original horizontal motion. The fluid particles, which move in
the direction of global rotation, are deflected outwards in radius
by the Coriolis force. Fluid particles moving in the retrograde
direction are deflected inwards. The product $u_r u_\phi$ is
positive in both cases. Note that the influence of the Coriolis
force on the original radial motion produces a negative correlation
$\langle u_ru_\phi\rangle < 0$ (Fig.~\ref{f2}). The sense of radial
angular momentum transport is controlled by the turbulence
anisotropy. Angular momentum is transported inward if radial mixing
prevails and outward for predominantly horizontal mixing (R\"udiger
\cite{R89}). In the present case, the anisotropy of horizontal
type is produced by the stable stratification. It may be noticed
that we do not find a meridional $\Lambda$-effect. The meridional
flux of angular momentum is of higher order in ${N^*}^{-2}$ compared
to the radial flux. This is a consequence of small anisotropy in the
horizontal plane. Therefore, mixing in stellar radiation zones does
not produce latitudinal differential rotation.

In a star during its spinning-down, turbulent viscosity and $\Lambda$-effect
both transport angular momentum outward. If the turbulent viscosity
in the radial direction is estimated as $\nu_{_\mathrm{T}} \approx
\tau\langle u_r^2\rangle$, the ratio of angular momentum fluxes
produced by the viscosity and the $\Lambda$-effect can be estimated
as
\begin{eqnarray}
    \frac{Q^\nu_{r\phi}}{Q^\Lambda_{r\phi}} &=&
    \frac{-\nu_{_\mathrm{T}}r\frac{\partial\Omega}{\partial
    r}\sin\theta}{\langle u_r u_\phi\rangle} =
    q \left(\frac{\Omega}{N}\right)^2 \ll 1,
    \nonumber \\
    q &=& -\frac{r}{\Omega}\frac{\mathrm{d}\Omega}{\mathrm{d}r} ,
    \label{25}
\end{eqnarray}
where Eq.~(\ref{24}) was used. For the upper radiation zone of the
Sun, $\Omega^2/N^2 \sim 10^{-5}$ (see Fig.~1 in Kitchatinov \&
R\"udiger \cite{KR08}). For faster rotating young stars the ratio is
larger, but still well below unity. Therefore, the angular momentum transport by
the $\Lambda$-effect is much more efficient compared to the effect
of the eddy viscosity.

Radial transport of chemical species is slow because it is only due
to weak radial mixing and it is further reduced by relatively
intensive horizontal mixing (Vincent, Michaud \& Meneguzzi
\cite{VMM96}; Michaud \& Zahn \cite{MZ98}). This may explain why the
characteristic times of rotational coupling between the core and the
convective envelope (\lsim 100 Myr; Hartmann \& Noyes \cite{HN87};
Denissenkov et al. \cite{Dea10}) in solar-type stars is shorter than
the characteristic time ($\sim$ 1 Gyr; Skumanich \cite{S72};
Mel\'endez et al., \cite{Mea10}; Baumann et al.\ \cite{Bea10}) of
Lithium depletion.

Papaloizou \& Pringle (\cite{PP78}) found that differential rotation
in radiative cores is unstable to $r$-modes or vortices that are
global in latitude but of small scale in radius. For the estimates
below we assume that the turbulent motions are global in the horizontal
directions and that radial eddy diffusion of chemicals is estimated
as
\begin{equation}
    \chi_{_\mathrm{T}} \approx \tau\langle u_r^2\rangle \approx
    \tau\langle u^2\rangle^{(0)} \frac{{\Omega^*}^2}{2{N^*}^4} \approx
    \frac{2R^2\Omega^2}{\tau^3N^4} ,
    \label{26}
\end{equation}
where again Eq.~(\ref{24}) was used. Lithium in young solar-type
stars is depleted about ten times in the first billion years of
their main-sequence life (Mel\'endez et al.\ \cite{Mea10}). Lithium
has to be transported over a short distance ($\sim 0.1 R$) below the
convection zone in order to be destroyed (see Fig.~1 of R\"udiger \&
Pipin \cite{RP01}). This leads to an estimated fractional Lithium
abundance (relative to the primordial abundance) at the age of 1~Gyr
\begin{equation}
    \mathrm{exp}\left(-\frac{10^9\ \mathrm{yr} \times\chi_{_\mathrm{T}}}{10^{-2}
    R^2}\right) \approx 0.1 .
    \label{27}
\end{equation}
Taking the value of $R \approx 5\times 10^{10}$~cm for the radius of
the base of (solar) convection zone, we get from Eq.~(\ref{27})
\begin{equation}
    \chi_{_\mathrm{T}} \approx 2\times 10^3\ \mathrm{cm}^2\
    \mathrm{s}^{-1} .
    \label{28}
\end{equation}
Equations~(\ref{26}) and (\ref{28}) lead to an estimate of the
characteristic eddy turnover time of
\begin{equation}
    \tau \approx 10^7\ \mathrm{s}
    \label{29}
\end{equation}
for a young star rotating ten times faster than the Sun. With this
value of $\tau$, an estimate for the characteristic time of
core-envelope rotational coupling due to the $\Lambda$-effect can be
obtained,
\begin{equation}
    T_\Omega \approx \frac{R^2\Omega}{Q^\Lambda_{r\phi}} \approx
    \tau^3N^2 \approx 60\ \mathrm{Myr} ,
    \label{30}
\end{equation}
where again Eq.~(\ref{24}) for the non-diffusive flux of angular
momentum $Q^\Lambda_{r\phi} = \langle u_r u_\phi\rangle$ was used
together with the estimate $N \approx 10^{-3}\, \mathrm{s}^{-1}$
for radiation zones.

The estimates suggest the following scenario for angular momentum
and chemical species transport in stellar radiation zones. Angular
momentum loss by a stellar wind induces differential rotation with
an angular velocity decreasing outwards. The differential rotation
is hydrodynamically unstable and this instability produces turbulent
mixing. The stable stratification makes the turbulence highly
anisotropic with very small radial velocities. The eddy diffusion in
radius lead to a slow decrease of the Lithium abundance with time.
Angular momentum is transported not by the eddy viscosity but by the
$\Lambda$-effect due to the correlation of azimuthal and radial
turbulent motions induced by the Coriolis force. The non-viscous
outward transport provides relatively fast rotational coupling
between the radiative core and the convective envelope.

Other possibilities for rotational coupling discussed in the
literature include the angular momentum transport by gravity waves
(Charbonnel \& Talon \cite{CT05}) and by magnetic stress
(Charbonneau \& MacGregor \cite{CM93}; R\"udiger \& Kitchatinov
\cite{RK96}; Denissenkov \cite{D10}). Further studies are necessary
to decide which of the possibilities are most relevant.
 \acknowledgements
LLK is thankful to NORDITA for hospitality and support. This work
was supported by the Russian Foundation for Basic Research
(projects 10-02-00148, 10-02-00391) and the European Research Council
under the AstroDyn Research Project 227952.
We acknowledge the allocation of computing resources provided by the
Swedish National Allocations Committee at the Center for
Parallel Computers at the Royal Institute of Technology in
Stockholm and the National Supercomputer Centers in Link\"oping.

\end{document}